\documentclass[12pt]{article}
\def\lsim{\mathrel{\rlap {\raise.5ex\hbox{$ < $}}
{\lower.5ex\hbox{$\sim$}}}}
\def\gsim{\mathrel{\rlap {\raise.5ex\hbox{$ > $}}
{\lower.5ex\hbox{$\sim$}}}} 
\def\sqr#1#2{{\vcenter{\vbox{\hrule height.#2pt
        \hbox{\vrule width.#2pt height#1pt \kern#1pt
           \vrule width.#2pt}
        \hrule height.#2pt}}}}

\def\lsim{{\displaystyle
{{\raise-8pt\hbox{$ <$}}
\atop{\raise5pt\hbox{$\sim$}}}}}
\def\gsim{{\displaystyle
{{\raise-8pt\hbox{$ >$}}
\atop{\raise5pt\hbox{$\sim$}}}}}
\def\slsim{{\displaystyle
{{\raise-8pt\hbox{$\scriptstyle <$}}
\atop{\raise5pt\hbox{$\scriptstyle \sim$}}}}}
\def\sgsim{{\displaystyle
{{\raise-8pt\hbox{$\scriptstyle  >$}}
\atop{\raise5pt\hbox{$\scriptstyle \sim$}}}}}
\newskip\humongous \humongous=0pt plus 1000pt minus 1000pt

\newcommand{\sumpf}[0]{\sum_{(H^{\rm f},G^{\rm f})}\! \! \! \!
{\raise
4pt
\hbox{$'$}}\,}

\newcommand{\sump}[0]{\sum_{(H,G)}\! \! {\raise 4pt \hbox{$'$}}\,}

\def\bs{\begin{subequations}}
\def\es{\end{subequations}}
\catcode`\@=11
\newcount\hour
\newcount\minute
\newtoks\amorpm
\hour=\time\divide\hour by60
\minute=\time{\multiply\hour by60 \global\advance\minute by-\hour}
\edef\standardtime{{\ifnum\hour<12 \global\amorpm={am}%
        \else\global\amorpm={pm}\advance\hour by-12 \fi
        \ifnum\hour=0 \hour=12 \fi
        \number\hour:\ifnum\minute<10 0\fi\number\minute\the\amorpm}}
\edef\militarytime{\number\hour:\ifnum\minute<10 0\fi\number\minute}
\def\draftlabel#1{{\@bsphack\if@filesw {\let\thepage\relax
   \xdef\@gtempa{\write\@auxout{\string
      \newlabel{#1}{{\@currentlabel}{\thepage}}}}}\@gtempa
   \if@nobreak \ifvmode\nobreak\fi\fi\fi\@esphack}
        \gdef\@eqnlabel{#1}}
\def\@eqnlabel{}
\def\@vacuum{}
\def\draftmarginnote#1{\marginpar{\raggedright\scriptsize\tt#1}}
\def\draft{\oddsidemargin -.2truein
        \def\@oddfoot{\sl preliminary draft \hfil
        \rm\thepage\hfil\sl\today\quad\militarytime}
        \let\@evenfoot\@oddfoot \overfullrule 3pt
        \let\label=\draftlabel
        \let\marginnote=\draftmarginnote
   \def\@eqnnum{(\theequation)\rlap{\kern\marginparsep\tt\@eqnlabel}%
\global\let\@eqnlabel\@vacuum}  }
\def\subequations{\refstepcounter{equation}%
  \edef\@savedequation{\the\c@equation}%
  \@stequation=\expandafter{\theequation}
  \edef\@savedtheequation{\the\@stequation}
  \edef\oldtheequation{\theequation}%
  \setcounter{equation}{0}%
  \def\theequation{\oldtheequation\alph{equation}}}

\def\endsubequations{\setcounter{equation}{\@savedequation}%
  \@stequation=\expandafter{\@savedtheequation}%
  \edef\theequation{\the\@stequation}\global\@ignoretrue
  \vspace*{-12pt} \\}
\def\bs{\begin{subequations}}
\def\es{\end{subequations}}
\relax
\def\Im{\,{\rm Im}\, }
\def\Re{\,{\rm Re}\, }

\def\thefootnote{\fnsymbol{footnote}}
\def\be{\begin{equation}}
\def\ee{\end{equation}}
\def\ba{\begin{eqnarray}}
\def\ea{\end{eqnarray}}

\def\th{\vartheta}

\newcommand{\ar}[2]{{#1\atopwithdelims[]#2}}

\def\ee{\end{equation}}
\def\bea{\begin{eqnarray}}
\def\eea{\end{eqnarray}}
\def\nn{\nonumber}

\def\np#1#2#3{Nucl. Phys. {\bf{B#1}} (#2) #3}
\def\pl#1#2#3{Phys. Lett. {\bf{B#1}} (#2) #3}

\def\pr#1#2#3{Phys. Rev. {\bf{D#1}} (#2) #3}

\newcommand{\uarrw}[0]{\mathrel{
{\raise.5ex\vbox{\hrule width 1cm}\hskip-6pt\rightarrow}}}
\def\thebibliography#1{%
\vskip 0.5cm \centerline{\bf References}
\list{%
[\arabic{enumi}]}{\settowidth\labelwidth{[#1]}
\leftmargin\labelwidth
\advance\leftmargin\labelsep
\usecounter{enumi}}
\def\newblock{\hskip .11em plus .33em minus .07em}
\sloppy\clubpenalty4000\widowpenalty4000
\sfcode`\.=1000\relax}

\renewcommand{\theequation}{\arabic{section}.\arabic{equation}}

\renewcommand{\section}{\setcounter{equation}{0}\@startsection%
{section}{1}{0mm}{-\baselineskip}{0.5\baselineskip}%
{\normalfont\normalsize\bfseries}}

\renewcommand{\subsection}{\@startsection%
{subsection}{2}{0mm}{-\baselineskip}{0.5\baselineskip}%
{\normalfont\normalsize\slshape}}
\begin{document}
\renewcommand{\theequation}{\arabic{section}.\arabic{equation}}
\begin{titlepage}
\begin{flushright}
Bicocca-FT-00-06,\\
hep-th/0005124 
\end{flushright}
\begin{centering}
\vspace{1.0in}
\boldmath
{\bf \large On the strong coupling behavior of heterotic 
and type I orbifolds$^\dagger$}
\\
\unboldmath
\vspace{1.7 cm}
{\bf Andrea Gregori$^1$} \\
\medskip
\vspace{.4in}
{\it  Dipartimento di Fisica, Universit{\`a} di Milano--Bicocca},\\
{\it and}\\
{\it  INFN, Sezione di Milano, Italy}\\
\vspace{2.5cm}
{\bf Abstract}\\
\vspace{.1in}
We study the bulk effective theory of a class of orbifolds of
the type IIB string with D5-branes, compactified to four dimensions. 
These constructions are connected, in a region of their moduli space, to 
some orbifolds of the type I and heterotic string. 
We compare the effective actions through the 
coupling of the $R^2$ term, and we argue that
these orbifolds provide non-perturbative deformations of the latter, 
in which the gauge group is entirely non-perturbative.  
\end{centering}
\vspace{2cm}

\hrule width 6.7cm
$^\dagger$\  Research partially supported by the EEC under the contract\\
TMR-ERBFMRX-CT96-0045.\\
\\
$^1$e-mail: agregori@pcteor.mi.infn.it

\end{titlepage}
\newpage
\setcounter{footnote}{0}
\renewcommand{\thefootnote}{\arabic{footnote}}

\setcounter{section}{1}
\section*{\normalsize{\bf 1. Introduction}}

In Ref. \cite{gmono} we analyzed a class of orbifolds of the type IIB string,
in the presence of parallel D5-branes, 
with partial, spontaneous breaking of the supersymmetry to ${\cal N}_4=2$.
We discussed how, in a region of the moduli space, they are
connected to the type I and heterotic string, 
arguing that these constructions provide
a possible non-perturbative deformation of type I and heterotic orbifold 
constructions.
These models were obtained by considering
freely acting orbifolds of the type IIB string projected with 
$I_{(7,8,9,10)} \Omega$, the world-sheet parity
times the reflection along four compact coordinates, and then looking at the 
theory from the bulk point of view. In the compactification limit of
vanishing size of the coordinates $x_7$, $x_8$, $x_9$, $x_{10}$,
these constructions could be matched with the ordinary
type I orientifold construction,  obtained by
projecting the type IIB string with 
$\tilde{\Omega} \equiv I_{(7,8,9,10)} \Omega$. In the following, 
with an abuse of language, we will refer to these models as to
the $\tilde{\Omega}$ constructions. 
Owing to the free action of the orbifold projection,
at a generic point in the moduli space the corresponding ordinary, 
type I construction did not have a D5-branes sector,
appearing only in a corner of the moduli space.
In the $\tilde{\Omega}$ construction, where D9- and D5- branes appear 
exchanged, it was the D9-branes sector to be missing
at a generic point in the moduli space: the open string sector
was therefore entirely provided by strings ending on D5-branes.
By choosing a configuration of branes and orientifold planes
such as to provide a local cancellation of the tadpoles, it was possible,
thanks to the existence of directions transverse to the branes, to 
look at the bulk theory in the limit of large volume of the transverse space,
$V_{(4)} \to \infty$. As discussed in Ref \cite{gmono}, 
assuming the bulk point of view amounts to performing a translation
resulting into an half-integer shift of the winding numbers corresponding 
to the transverse coordinates: in the $V_{(4)} \to \infty$ limit
the action of the D5-branes and of the
associated O5 planes is ``trivialized''. 
From  the bulk point of view, the breaking of supersymmetry due to the 
$\tilde{\Omega}$ projection is seen as  ``spontaneous'',  corresponding, 
in the effective theory, to a non-perturbative super-Higgs mechanism.
This decompactification limit exists only for a configuration of the 
D5-branes and orientifold planes such that there is a local
cancellation of the Ramond-Ramond charge. In the case of D5-branes,
the gauge group must be broken to $U(1)^{16}$. 
More generically, a decompactification
limit may exist also for larger gauge groups, when they  correspond to
Dp-branes configurations, p $<$ 9, with a local cancellation
of the RR charge with the appropriate orientifold planes.
For $SO(32)$, this limit does not exist \cite{pw}.

In this work, we continue the analysis of Ref. \cite{gmono}, by
considering the case of non-freely acting orbifolds.
The breaking of supersymmetry is in this case non-spontaneous,
and it is not possible to restore a larger amount of supersymmetry
by going to special limits in the moduli space. On the other hand,
these orbifolds have fixed points, that provide 
${\cal N}_4=2$ vector and hypermultiplets. From the heterotic point of view, 
the extension of the gauge group is non-perturbative.
The examples we consider actually include a construction that matches,
in a corner of the moduli space, a phase of the
type I ``$U(16) \times U(16)$'', ${\cal N}_4=2$ orbifold \cite{sagn,gp},
in which appropriate Wilson lines break the gauge group to 
$U(1)^{16} \times U(1)^{16}$. 

By comparing the expression of the coupling of the $R^2$ term 
in the four-dimensional effective actions of the
$\tilde{\Omega}$ constructions, considered in different limits of the 
moduli space,
with those of the related heterotic and type I theories,
we conclude  that these models can be interpreted
as ``semi-freely acting orbifolds'', in which the non-spontaneous breaking
of supersymmetry is due to the interference of a ``freely acting''
projection, corresponding to the ``$\tilde{\Omega}$'' construction,
and a non-freely acting one, the $Z_2$ orbifold projection.
The different nature of these two projections is reflected in the
asymmetry between the D9- and D5-branes sectors. 
Indeed, at a generic point in the moduli space,
the ``D9'' sector is massive, and the only gauge group 
is provided by the D5 branes, that, in  the $\tilde{\Omega}$ orbifolds,
appear as space-fulfilling, D9 branes.

The paper is organized as follows:
in Section 2 we discuss the $\tilde{\Omega}$ constructions. 
In Section 3 we describe their type I duals, while 
the general properties of the heterotic duals are
discussed in Section 4.
In Section 5 we compare these theories through the analysis of the
corrections to the $R^2$ term. As in Ref. \cite{gmono},
we compare here the heterotic and type I computations
with the result obtained on the ``$\tilde{\Omega}$'' side in the two limits
of small and large volume, $V_{(4)}$, of the space transverse to the D5-branes.
For small $V_{(4)}$, the construction matches the 
ordinary, type I orientifolds, while in the opposite limit,
there is a non-trivial behavior, due to the suppression  of the 
D5-branes effects. 

We devote Section 6 to further comments and conclusions.

\noindent

\vskip 0.3cm
\setcounter{section}{2}
\setcounter{equation}{0}
\section*{\normalsize{\bf 2. The $\tilde{\Omega}$ orientifolds }}

We start by reviewing here the action of the projection $\tilde{\Omega}$
on the type IIB string. We consider directly the situation that
will be of interest for us in the following, namely the case in which
six coordinates are compactified on a product of circles. 
The ordinary orientifold construction is obtained by modding
the type IIB string by the world-sheet parity, $\Omega$. In less than
ten dimensions, it is possible to consider, instead of 
$\Omega$, the operation $\tilde{\Omega}$, defined as the world-sheet parity 
times the target space reflection, $I_x: \; x \to -x$, 
along an even number of compact coordinates $x$. When we mod the type IIB
string with $\tilde{\Omega} \equiv \Omega \times I_{(7,8,9,10)}$,
we obtain an orientifold in which the O9 plane is
replaced by O5 planes and the D9-branes by D5-branes.
It is still possible to have a local tadpole cancellation, simply
by putting each one of the sixteen D5-branes, together with
its mirror, on one of the sixteen O5 orientifold planes.
In this configuration, the initial ${\cal N}_4=8$ supersymmetry 
of the type IIB string is reduced to ${\cal N}_4=4$, and the
massless spectrum is the same as that of the ordinary type I string
in four dimensions, with the gauge group broken to $U(1)^{16}$.
The partition function of this orientifold corresponds
to that of the type I string, with four coordinates T-dualized:
their contribution appears therefore with the inverse of the radius and with
winding numbers instead of momenta.
As we remarked in Ref. \cite{gmono}, owing to the breaking
of T-duality along the coordinates transverse to the D5-branes,
in the type IIB string with $\tilde{\Omega}$ it is possible
to assume a point of view from which
the partial breaking of supersymmetry appears as ``spontaneous''. 
This corresponds to looking at the system from the ``bulk''.
The partition function of this ``displaced'' theory is easily
obtained from the ordinary orientifold partition function by
performing a ``translation'' toward the center of the 
segments in the coordinates transverse to the D(5)-branes.
This operation amounts to a shift in the windings in the
Klein bottle and in the open string diagrams. In the 
``transverse channel'' this appears instead as a rescaling 
of the momenta by an amount proportional to the fraction
of the segment it has been translated. 
In the closed string sector, the Torus, being by itself
invariant under translations, is not affected by this operation.
From the point of view of this picture, the theory coincides with the ordinary
type I orientifold in the $V_{(4)} \to \infty$ limit.
Indeed, in this limit the most correct description is obtained when 
the transverse coordinates 
are T-dualized, giving the ordinary type I construction.
In the $V_{(4)} \to \infty$ limit there is instead
an approximate restoration  of the initial supersymmetry, 
with the disappearance of the D5-branes sector.
This phenomenon cannot be observed in the ten dimensional constructions, 
because the orientifold projection produces there D9-branes,
that fulfill the space: there are therefore no transverse directions 
along which to move and go  ``far from the branes''.
This motion can be observed only when, after compactification,
 appropriate Wilson lines are switched on: these can be interpreted
as the un-freezing of ``transverse'' directions.

When we add a further $Z_2$ orbifold projection, acting as a
twist along $x_7$, $x_8$, $x_9$, $x_{10}$, supersymmetry
is further broken to ${\cal N}_4=2$. In Ref. \cite{gmono},
the twist was coupled to a translation along $x_4$ and/or $x_5$,
inside the untwisted two-torus. The further breaking of supersymmetry
was therefore spontaneous, and, from the bulk point of view, 
the theory possessed a spontaneously broken ${\cal N}_4=8$ supersymmetry.
We consider now instead the case in which there is no translation.
This $Z_2 \equiv Z^{\rm o}_2$ orbifold possesses then sixteen fixed points.
The combined action of the $Z^{\rm o}_2$ and $\tilde{\Omega}$
projections generates also a D9-branes sector, absent in the previous case.
This sector provides, at the Abelian point, sixteen vector multiplets.
The analysis of the decompactification limit in this ${\cal N}_4=2$
theory is not as simple  as it was for the ${\cal N}_4=4$ case:
strictly speaking, owing to the simultaneous presence of both D5- and
D9-branes, this limit is always singular. 
Even though in the $V_{(4)} \to \infty$ limit
the D5-branes sector is ``trivialized'', in the sense above 
described for the ${\cal N}_4=4$ case,
the breaking of supersymmetry is in this case non-spontaneous: 
as we will see, there is no limit in the
moduli space in which the ${\cal N}_4=2$, D9-branes sector, disappears,
and the theory is therefore always in the phase of reduced,
${\cal N}_4=2$ supersymmetry.

The $S^1 \big/ Z_2$ orbifolds possess a symmetry that exchanges the states
in the twisted sector, constructed on the twist fields $\sigma_+$,
$\sigma_-$, and projects the momenta in the states $V_{mn}$ of the 
untwisted sector (see Ref. \cite{dvv}):
\be
{\cal D} \, : ~~~~(\sigma_+,\sigma_-,V_{mn}) ~~ \to ~~
(\sigma_-,\sigma_+, (-1)^m \, V_{mn}) \, .
\label{D}
\ee
In our $Z^{\rm o}_2$ orbifold there are four directions 
possessing this symmetry:
by coupling the ${\cal D}$-operation on the twisted coordinates
to a translation in the two-torus, it is therefore possible to construct
orbifolds with a reduced number of twisted states \cite{gkp}--\cite{gkr}.
This operation commutes with the supersymmetry-breaking projection.
By projecting with one such operation\footnote{We refer the reader to 
\cite{gkp}--\cite{gkr} for a
detailed discussion of this operation.}, the number of fixed points 
is reduced by half; with two of them it is reduced to 
a quarter of the initial one.
This is the maximal reduction, because there are only two independent 
directions in the two-torus, along which it is possible to act with the
associated translation.
Starting from the model discussed above, with
sixteen hypermultiplets originating from the twisted sector,
and a corresponding D9-branes sector with sixteen vector multiplets, 
it is possible to construct, via ${\cal D}$-projections,
models with respectively eight and four of them.
All these models differ only in the $Z_2$-twisted and the D9-branes sector, 
the rest of the spectrum being the same (see also Section 3).
 
\noindent

\vskip 0.3cm
\setcounter{section}{3}
\setcounter{equation}{0}
\section*{\normalsize{\bf 3. The type I duals}}

The type I dual of the orbifold with sixteen fixed points is the
model with maximal
gauge group $U(16) \times U(16)$, presented in \cite{sagn,gp}, which is
constructed as an orientifold of the type IIB string compactified
on $T^2 \times T^4 \big/ Z_2$. 
The second $U(16)$ factor, that originates from the D5-branes sector,
corresponds, on the heterotic side, to the contribution of
small instantons \cite{w}, and is entirely non-perturbative.
The gauge group $U(16) \times U(16)$ is obtained when
all the D5-branes sit at an orbifold fixed point \cite{gp}.
The open string massless spectrum contains vector multiplets in the adjoint 
representation of $U(16) \times U(16)$ and hypermultiplets
in the $({\bf 120}, {\bf 1})$ and $({\bf 1}, {\bf 120})$,
antisymmetric representations, and in the $({\bf 16},{\bf 16})$,
bifundamental representation.  
The massless spectrum of this model includes then 
the states of the closed string sector, namely
the sixteen hypermultiplets originating from the orbifold twisted sector
and the usual three vector and four hypermultiplets of the compactification.
By introducing appropriate Wilson lines and separating the D5-branes
from each other, it is possible to break the gauge group
$U(16) \times U(16)$ to $U(1)^{16} \times U(1)^{16}$. 
At this point, the second $U(1)^{16}$ factor can be shown
to be anomalous \cite{6danom}. The anomaly cancellation mechanism
gives also a mass to the hypermultiplets originating from
the twisted sector. Due to the presence of the Wilson lines 
breaking to the Abelian subgroup the first factor,
only sixteen hypermultiplets remain.

The type I duals of the orbifolds with reduced number of fixed points
are easily constructed as orientifolds of the type IIB string, 
compactified on $T^2 \times T^4 \big/ Z_2$ and projected with 
${\cal D}$-operations. 
Choosing $(-1)^m$ for the action on $T^2$ associated to the 
${\cal D}$-projection leads to a shift that 
lifts the mass of half of the  states of the closed string twisted
sector by projecting their momenta. This operation 
affects therefore  the torus amplitude and the D5-branes contribution
in the Annulus and M\"{o}bius strip amplitudes.
The partition function of this model is 
given by half the partition function of the ordinary orbifold plus half
the partition function of the ``momentum breaking'' freely acting orbifold
of Ref. \cite{adds}.
The tadpole cancellation conditions for the D9-branes gauge group are the same;
therefore, the part that corresponds to the perturbative heterotic
gauge group is not affected.
The tadpole conditions for the D5-branes instead are changed:
now they are solved by putting half of the usual number of D5-branes.
The rank of the non-perturbative gauge group is reduced by half 
per each such ${\cal D}$-projection. At the Abelian point the gauge group 
is therefore $U(1)^{16}_9 \times U(1)^{16 \big/ 2^{n_{\cal D}}}_5$, with 
$n_{\cal D}$ the number of ${\cal D}$-projections.

\noindent

\vskip 0.3cm
\setcounter{section}{4}
\setcounter{equation}{0}
\section*{\normalsize{\bf 4. The heterotic duals}}

The heterotic dual of the $U(16) \times U(16)$ type I orientifold,
with gauge group broken to $SU(16)$, was explicitly constructed
in Ref. \cite{ap2}, by starting from the ${\cal N}_4=4$
heterotic string with gauge group $SO(32)$.
As discussed in Ref. \cite{6danom}, this construction is in the same
moduli space of the ${\cal N}_4=2$ heterotic theory 
derived from the ${\cal N}_4=4$ theory with gauge group $E_8 \times E_8$,
with an embedding of the spin connection into the gauge group realized
by 12 instantons in each of two  $E_8$'s \cite{dmw}. 
This second construction can be realized as a
$T^2 \times T^4 \big/ Z_2$ orbifold
of the ${\cal N}_4=4$ heterotic string compactified on $T^6$ \cite{fi}.
This is the construction we will consider here.
From this, we will derive also the duals of the orbifolds
with a reduced number of fixed points.

We start therefore with the dual of the orbifold with sixteen
hypermultiplets in the twisted sector: it corresponds to the
maximal rank heterotic construction, in which 
the gauge group embedding of the spin connection is realized
as follows.
The $Z_2$ orbifold projection acts symmetrically
on the left and right movers of $T^4$ and has a $c=(0,6)$ embedding
in the currents. This means that if the $c=(0,16)$ block
corresponding to the currents is described in terms of 32 free
fermion degrees of freedom, the $Z_2$ projection acts
as $-1$ on 12 of these fermions.
An embedding of this form is not necessarily required by modular invariance,
that would be satisfied also by a $c=(0,2)$ embedding: 
our choice is dictated by the fact that
we want the $Z_2$-twisted sector to be massless.
However, this is not enough. If we count the number of
hypermultiplets of the $Z_2$-twisted sector, we see that they exceed
the number of orbifold fixed points.
This is due to the fact that, owing to the embedding of the spin connection
into the gauge group, on the heterotic side these hypermultiplets
are charged under the gauge group. 
In order to reproduce the ``type I'' duals, we
must break to $U(1)$'s the factor of the gauge group under which these 
hypermultiplets are charged. This is realized through the introduction
of further Wilson lines. Now, if we count the number of such hypermultiplets
at the $U(1)$ point, we see that they are half of those of the type I. 
A way of seeing this is by counting the scalars of these hypermultiplets:
we have a factor sixteen coming from the sixteen fixed points of the
$T^4 \big/ Z_2 $ orbifold, times a factor two, 
the multiplicity of a $spin (4)$ corresponding to  
the four twisted left-moving fermions, of which the GSO projection 
forces to choose one of the two chiralities.
Since the gauge group is $U(1)^{16}$, there are no
additional multiplicities coming from the part embedded in the gauge group,
as we wanted. We have therefore half of the scalars of sixteen
hypermultiplets.
In order to obtain the correct number, we must introduce a further,
discrete Wilson line Y, whose action can be easily described in terms
of free fermions: it is introduced by a set of four complex fermions,
two of them intersecting the $Z_2$ orbifold
projection on the currents. Besides this, $Y$
acts also as a shift in one of the twisted coordinates of the 
compact space. 
To be concrete, let's start by the case in which, at the ${\cal N}_4=4$
level, the gauge group of the currents is broken to $SO(16) \times  SO(16)$.
The $Z_2$ orbifold projection is then
embedded in an $SO(8)$ subgroup of the first $SO(16)$ and 
an $SO(4)$ subgroup of the second $SO(16)$. The Wilson line Y
picks then a factor $SO(4)$ in each $SO(16)$: in the first $SO(16)$ factor
the $SO(4)$ embedding of Y is a subgroup of the $SO(8)$ of the orbifold
projection. The gauge group is in this case broken to 
$\left[ SO(8) \times \left( SO(4)_Y \times SO(4) \right)_{Z_2} \right]
\times \left[ SO(8) \times SO(4)_Y \times SO(4)_{Z_2} \right] $:
Y~plays therefore the role of symmetrizing the $Z_2$ orbifold 
projection in the two $SO(16)$ factors. 
Besides the $Z_2$~-twisted sector, there is now also the massless
$Z_2 \times Y$~-twisted sector, that provides the other half of the
states we needed in order to form the sixteen hypermultiplets of the 
twisted sector.

As in the dual orbifolds, the number of hypermultiplets
originating from the $Z_2$-twisted sector can be reduced by half
or to a quarter of the initial one by modding out the twisted states
with ${\cal D}$-projections. Such projections act on $T^2$ as translations.

In all these models, at the $U(1)^{16}$
point,  the spectrum of the massless states
originating from the $c=(0,16)$ ``currents'' contains 
only sixteen vector multiplets.  
In order to break completely the gauge group, it is in fact necessary
to introduce Wilson lines that eliminate all the hypermultiplets:
the only remaining are those of the compact space (four), and those
of the orbifold twisted sector, whose number coincides with
the number of fixed points.

\noindent

\vskip 0.3cm
\setcounter{section}{5}
\setcounter{equation}{0}
\section*{\normalsize{\bf 5. Comparison of the models and
the effective strong coupling behavior}}

Before considering the $\tilde{\Omega}$ constructions, we
review the comparison of heterotic and type I strings, already
analyzed in previous works \cite{ap1,ap2}. We
first recall some basic aspects of this duality, which can be verified
by looking at the renormalization of what in Refs. \cite{gkp}--\cite{gk}
was called the ``regular'' gravitational term, a combination
of $R^2$ and $F^2$ amplitudes, that possesses the property of being 
an ``holomorphic'', smooth function of the moduli in all the string
constructions (see below).
We then consider the analogous corrections in the 
$\, \tilde{\Omega} \, $ constructions, both in the limit
$V_{(4)} \to 0$ and $V_{(4)} \to \infty$. In the first limit, where 
the $\tilde{\Omega}$ theory matches the type I string,
the correction coincides with that of the latter.
In the $V_{(4)} \to \infty$ limit, 
the D5-branes effects are instead suppressed,
and the model behaves as an ${\cal N}_4=2$, type I orbifold without D5-branes,
like the Scherk--Schwarz breaking model considered in Refs. \cite{adds,gk}.
From the point of view of the effective four dimensional theory,
matching the heterotic and type I string in a corner of the moduli space, 
this is a non-perturbative limit.

\subsection*{\sl The $R^2$ corrections}

As we said, in order to compare different string constructions, we must 
look at an amplitude regular in all the theories.
As it was discussed in Refs. \cite{gmono,gkp,gk},
the correct amplitude is obtained by projecting out of the $R^2$ amplitude
the terms that mix ``bulk'' and ``branes'' contribution.
This subtraction corresponds, on the heterotic string side,
to the term:
\be
\langle R^2 \rangle_{(\rm o)} =
\langle R^2 \rangle 
+ {1 \over 12} \langle P^2 \rangle_{(T^2)} 
+ {5  \over 48} \langle F^2  \rangle_{\rm gauge}  \, . 
\label{rff}
\ee
The second term in the r.h.s.
corresponds to the amplitude of the $U(1)^2$ of the two-torus,
computed as in \cite{gkp}; the third is proportional
to the gauge amplitude of the currents. 
The subtraction of the torus amplitude is also necessary,
because it has singularities absent in the dual string constructions.   
On the type I side, the one-loop contribution to the $R^2$ term
is proportional to an index \cite{ser}:
this subtraction amounts therefore to a simple rescaling of the
beta-function coefficient. In order to compare with the heterotic side,
this rescaling may be necessary, because we must subtract the contribution
of the D5-branes states, non-perturbative on the heterotic side.
In the present case (at the Abelian point), 
all the extra gauge multiplets are anomalous and
there is no such a non-perturbative extension.
On the $\tilde{\Omega}$ duals, the same considerations are valid,
with the role of D9- and D5-branes exchanged.
As we learned in \cite{gmono},
in the $V_{(4)} \to \infty$ limit, the D5-branes effects are suppressed
and we expect the effective coupling of this amplitude
to depend only on $U$, the modulus associated
to the complex structure of the two-torus
around which the D5-branes are wrapped, and on a ``tree level'' contribution,
proportional to the vacuum expectation value of a field,
$\Im \tilde{S}^{\prime}$, corresponding to the coupling of the 
D9-branes sector, always present. 

\subsection{\sl The type I amplitudes}

Owing to extended supersymmetry, the corrections to the $R^2$ and $F^2$ 
amplitudes receive a contribution only from the tree and one loop levels.
Putting together tree level and one-loop contributions, we get the
following expression for the effective coupling: 
\be
{16 \, \pi^2 \over g^2} = 16 \, \pi^2  \,\Im S \, +
16 \, \pi^2  \, \Im S^{\prime} \, +
\Delta^{N^{(\rm t)}}(U) +
b \log M_P \big/ \sqrt{p^2} \, .
\label{Ig}
\ee
The first two terms on the r.h.s. give the tree-level contribution.
The fields  $S$ and $S^{\prime}$, whose  imaginary parts
parametrize the gauge coupling in the D9-branes
and the D5-branes sector respectively, are 
defined as in Refs. \cite{par}--\cite{ap2}:
\ba
\Im S & = & {\rm e}^{-\phi_4} G^{1 / 4} \omega^2 \, ,  
\label{s} \\
&& \nonumber \\
\Im S^{\prime} & = & {\rm e}^{-\phi_4} G^{1 / 4} \omega^{-2} \, ,  
\label{sp}
\ea
where $\phi_4$ is the type I dilaton of four dimensions,
$\sqrt{G}=R_4 R_5$ the volume of the untwisted two-torus and
$\omega^4$ the volume of the K3 ($\sim T^4 \big/ Z_2$).
The real parts, $\Re S$, $\Re S^{\prime}$, are the scalars
dual respectively to $B_{\mu \nu}$ and $B_{45}$.
The function $\Delta^{N^{(\rm t)}}(U)$ is, for $N^{(\rm t)}=16$
orbifold fixed points:
\be
\Delta^{16}(U) \, = \,
-2 \log \Im U \vert \eta (U) \vert^4 \, .
\label{dtu}
\ee
In the other cases, it coincides with the $\Delta^{N^{(\rm t)}}(U)$
term of the heterotic corrections given in Eqs. (\ref{delta8}), (\ref{delta4})
below. The beta-function coefficient $b$, that depends on
$N^{(\rm t)}$, is the same as in Eqs. (\ref{rIIB}) and (\ref{rhet}).

\subsection{\sl The ``$\, \tilde{\Omega} \, $'' amplitude in the 
limit $V_{(4)} \to \infty$}

According to the observations of Ref. \cite{gmono}, 
in the large-$V_{(4)}$ limit, the type IIB orbifold
with $\tilde{\Omega}$ orientifold projection, as seen from the bulk point
of view, behaves as an ${\cal N}_4=2$, type I orientifold without D9-branes.
Therefore, in the this limit,
we expect the correction to depend only on a modulus parametrizing the
coupling constant of the $\tilde{\Omega}$ D9-branes sector and
on the modulus $U$, given as before. We have then: 
\be
{16 \, \pi^2 \over g^2} \, \approx \,
16 \, \pi^2 \Im \tilde{S} \, + \,
\Delta^{N^{(\rm t)}=16}(U) \, + \,
b \log M_P \big/ \sqrt{p^2} \, ,
\label{rIIB}
\ee
where the function $\Delta^{N^{(\rm t)}}(U)$ is given as in Eq. (\ref{Ig})
and $\approx$ indicates that in Eq.(\ref{rIIB}) we are neglecting
terms suppressed with the volume of the transverse space, $V_{(4)}$.

Owing to the symmetry between the D9- and D5-branes
sectors of the type I construction, we could think to repeat the 
argument and perform a translation toward the bulk in the type I construction,
in order to observe the disappearence  of the D5-branes sector.
This is however not possible: this sector is produced by an orbifold
projection that we expect to preserve S-duality of its coupling.  
More precisely, even though the branes in the open string sector
are pushed at infinity, in the closed string sector the orbifold projection
still works, breaking supersymmetry to ${\cal N}_4=2$ at any point in the 
bulk, and the branes states we have apparently lost reappear
in the orbifold twisted sector. 
For what matters the effective coupling of the $R^2$ term, 
we may argue that the above property reflects into the invariance of 
the effective coupling under the 
inversion of the modulus parametrizing the effective coupling of the
D9-branes sector. This is in fact what we will find by matching,
in a region of the moduli space, this construction with the heterotic string. 
The first term in Eq. (\ref{rIIB}) can in fact be
seen as the $T \to \infty$ limit of: 
\be
\Delta^{N^{(\rm t)}=16}(T,U) \, = \,
-2 \log \Im T \vert \eta (T) \vert^4 \, ,
\label{dtu}
\ee
with the identification $T \sim \tilde{S}^{-1} \equiv S^{\prime}$.
The last term in the r.h.s. of Eq. (\ref{rIIB})
gives the infrared running in terms
of the Planck mass $M_P$ and a physical cut-off, $\sqrt{p^2}$;
the coefficient is $b= 6 + N^{(\rm t)} \big/ 4$,
where $N^{(\rm t)}$ is the number of the orbifold fixed points.

\subsection{\sl The heterotic amplitude}

On the heterotic side, there is a tree level, universal contribution.
The one-loop contribution is obtained by inserting in the
vacuum amplitude the operators corresponding to (\ref{rff}).
The left-moving part of any such operator is the helicity operator
$Q^2$, that leads to the saturation of the fermion zero modes.
The right moving parts are differential operators that, after
the saturation of the zero modes, act on various terms of the second 
helicity supertrace, $B_2$ \footnote{For a definition of helicity supertraces,
see for instance Ref. \cite{kbook}, and, for a computation in the context
of heterotic ${\cal N}_4=2$ orbifolds, Ref. \cite{kkprn}.}.
In order to compute these quantities, we need an explicit expression of the
partition function. 
Unfortunately, it is not possible to construct the heterotic dual at the 
$U(1)^{16}$ point by using only discrete, $Z_2$-Wilson lines.
This is a technical fact, due to the strong constraints imposed,
on the heterotic string, by modular invariance.
However, we can bypass
this difficulty by constructing the heterotic model
at an extended symmetry point, the $U(2)^8$ point.
Even though at this point the number of hypermultiplets of the
twisted sector is twice that of the type IIB orbifold,
the conclusions we will get are valid also at the $U(1)^{16}$
point. As we will see, in passing from the $U(2)^8$
to the $U(1)^{16}$ point, both the numbers of vector
and ``untwisted'' hypermultiplets are reduced, but their 
difference $N_V-N_H$ remains
the same. Universality properties (see Ref.\cite{kkprn}) then
tell us that the ``${\cal N}_4=2$'' sectors of the two orbifolds
are the same, and this is enough for our analysis,
because these are the only sectors that contribute to the helicity 
supertrace $B_2$. 
The partition function at the $U(2)^8$ point is:
\ba
Z_{\rm Het} & = &
{1 \over \Im \tau   | \eta|^4 } {1 \over 2}
\sum_{H,G}
Z_{6,22} \ar{H}{G} \nonumber \\
&& \times {1 \over 2} \sum_{a,b}{1\over \eta^4}~ (-)^{a+b+ab}
\vartheta \ar{a}{b}^2
\vartheta \ar{a+H}{b+G}
\vartheta \ar{a-H}{b-G} \, ,
\label{hH}
\ea
where $\{H,G \} \in \{0,1 \}$ parametrize the boundary conditions
introduced by the $Z_2$ orbifold projection
in the directions 1 and ${\bf {\tau}}$ of the world-sheet torus; 
the second line of (\ref{hH}) stands for the contribution of the 10 left-moving
world-sheet fermions $\psi^{\mu},\Psi^I$ and the ghosts
$\beta,\gamma$ of the super-reparametrization;
$Z_{6,22}\ar{H}{G}$ accounts for the $(6,6)$
compactified coordinates and the $c=(0,16 )$ conformal system, which
is described by 32 right-moving fermions, $\Psi_A$, $A=1,\ldots,32$:
\be
Z_{6,22}\ar{H}{G}=
{1\over 2^{4}}\, \sum_{\vec h, \vec g}~{1\over \eta^6 {\bar \eta}^6 }
\Gamma_{2,2} \ar{0}{0}
\,
\Gamma_{4,4} \ar{H \vert \vec h}{G \vert \vec g}\,
{\overline \Phi}\ar{H, \vec h}{G, \vec g} \, (-1)^{hg} \,
, \label{hH622}
\ee
with
\ba
\Phi \ar{H, \vec h}{G, \vec g} & = &
{1 \over 2} \sum_{\gamma,\delta}
{1 \over \eta^{16}} \, \times \left\{
\theta \ar{\gamma + h_1 + h}{\delta + g_1 +g}
\theta \ar{\gamma + h_1 - h}{\delta + g_1 -g}
\theta^2 \ar{\gamma + h_1}{\delta + g_1}  \right.
\nn \\
&& \nn \\
&& \left. ~~~~~ \times \,
\left( \theta \ar{\gamma + h_1 + h_2}{\delta +g_1 + g_2}
\theta \ar{\gamma +h_1 - h_2 - h_3}{\delta + g_1 - g_2 - g_3}
\theta \ar{\gamma + h_1 + h_2 - H}{\delta +g_1 + g_2 - G} \right. \right.
\nn \\
&& \nn \\
&& ~~~~~~~~~~~~~~~ \left. \left.
\theta \ar{\gamma +h_1 + h_2 + h_3 + H}{\delta + g_1 + g_2 + g_3 + G} \right)
\right\} \,  \nn \\
&& \nn \\
&& \times ~ \left\{
\theta \ar{\gamma + H + h}{\delta +G + g}
\theta \ar{\gamma + H - h}{\delta +G - g}
\theta^2 \ar{\gamma}{\delta} \right.
\nn \\
&& \nn \\
&& \left. ~~~~~ \times \, \left(
\theta \ar{\gamma + h_2}{\delta + g_2}
\theta \ar{\gamma - h_2 - h_3}{\delta - g_2 - g_3}
\theta \ar{\gamma + h_2 + H}{\delta + g_2 + G} \right. \right. \nn \\
&& \nn \\
&& \left. \left. ~~~~~~~~~~~~~~~~~~~
\theta \ar{\gamma + h_2 + h_3 - H}{\delta + g_2 + g_3 - G} \right)
\right\} \, .
\label{phi}
\ea
Here, $({\vec h}, {\vec g}) \equiv \{ (h,g),(h_1,g_1),(h_2,g_2),(h_3,g_3) \}$
indicates the projections introduced by the Wilson lines: Y$_1$, that
breaks $SO(32)$ to $SO(16) \times SO(16)$, Y, that symmetrizes
the action of $Z^{(H,G)}_2$ in the two $SO(16)$ factors, and
Y$_2$, Y$_3$, that break further the gauge group to 
$\left( SO(4) \times U(1) \times U(1) \right)^4 \cong U(2)^8$.   
In order to make Eq.(\ref{phi}) easily readable, we have separated
the two ``SO(16)'' blocks by braces $\{ \, \} \times \{ \, \}$.
Within these blocks, the characters corresponding to the
$U(1)^4$ factors are collected in the brackets $( \,)$. 
The contribution of the bosons of the compact space
is written in terms of shifted-twisted lattice sums:
$\Gamma_{2,2} \ar{0}{0}$ is the contribution of the two-torus,
and the arguments $\ar{0}{0}$ indicate that there is no shift of the 
momenta; $\Gamma_{4,4} \ar{H|\vec{h}}{G|\vec{g}}$ is the contribution
of the twisted $T^4$, where the first column in the argument 
indicates the twist, the second indicates the shifts produced by
the Wilson lines\footnote{For more details on twisted-shifted lattice sums,
we refer the reader to \cite{6auth,kkprn}.}.
The spectrum of the untwisted sector of this orbifold contains 
the vector multiplets of the $U(1)^2$ of the two-torus
and of the adjoint representation of the gauge group $U(2)^8$;
the hypermultiplets are: four from the twisted $T^4$
and sixteen in the $({\bf 2},{\bf 2})$ of
the four factors $SO(2) \times SO(2) \cong U(1) \times U(1)$.
In the twisted sector, there are 
$\left({1 \over 2}  \right) \times 16$ hypermultiplets in the
$({\bf 2},{\bf 1})$ and $\left({1 \over 2}  \right) \times 16$
in the $({\bf 1},{\bf 2})$ of $spin(4) \otimes spin (4)$.
When the $U(2)^8$ gauge group is broken to $U(1)^{16}$,
we obtain precisely sixteen hypermultiplets from the twisted sector.

As we said, the relevant quantity for the one-loop computation of the 
gravitational corrections is the second helicity supertrace, $B_2$. 
After a straightforward computation, we obtain:
\be
B_2 = {1 \over \overline{\eta}^{24}} 
\sump  \Gamma_{2,2} \ar{0}{0}
\overline{\Omega} \ar{H}{G} \, ,
\ee
where the prime indicates that the sum is taken only over the values
$(H,G)$$=\{(0,1)$, $(1,0)$, $(1,1) \}$. In our case,
the modular forms $\Omega \ar{H}{G}$ read:
\ba
\Omega \ar{0}{1}&=&{\hphantom{-}}
\frac{1}{2}
\left(\th_3^4+\th_4^4\right)
\th_3^8\, \th_4^8
\nonumber\\
\Omega \ar{1}{0}&=&-
\frac{1}{2}
\left(\th_2^4+\th_3^4\right)
\th_2^8\, \th_3^8
\label{Om88a}\\
\Omega \ar{1}{1}&=&{\hphantom{-}}
\frac{1}{2}
\left(\th_2^4-\th_4^4\right)
\th_2^8\, \th_4^8 \, .\nn
\ea
Acting on $B_2$ with the right-moving part of (\ref{rff}), 
$\overline{R}^2_{(\rm o)}$, we obtain:
\be
\overline{R}^2_{(\rm o)} B_2 \, = \,
{2 \over 3} \, \sump \Gamma_{2,2} \, \ar{0}{0} \, .
\label{rb}
\ee
After infrared-regularization \cite{infra}
and integration over the fundamental domain, we finally get the one loop 
contribution\footnote{Notice that $\Omega \ar{H}{G}$ is twice the 
$\Omega^{(0)} \ar{H}{G}$ part of the helicity supertrace in the $N_V=N_H$ 
freely acting orbifolds
considered in Refs. \cite{gkp,gk,gkp2}.},
that, summed to that of tree level, gives:
\be
{16 \, \pi^2 \over g^2} = 16 \, \pi^2 \Im S \, + \,
\Delta^{N^{(\rm t)}=16}(T,U) \, + \,
b \log M_P \big/ \sqrt{p^2} \, ,
\label{rhet}
\ee
where the function $\Delta^{N^{(\rm t)}=16}(T,U)$ is given as in
Eq. (\ref{dtu}), with $T$ and $U$ the moduli associated
respectively to the K\"{a}hler class and the complex structure of $T^2$.
In the orbifolds with a reduced number of fixed points,
the dependence on the moduli $T$ and $U$ is different.
In the case of half number of fixed points, i.e. with $N^{(\rm t)}=8$, 
the contribution of the two-torus,
instead of $\Gamma_{2,2} \ar{0}{0}$, is, both on the type IIB and
heterotic sides:
\be
{1 \over 2} \Gamma_{2,2} \ar{0}{0} + 
{1 \over 2} \Gamma_{2,2} \ar{H}{G} \, ,
\ee
where the arguments indicate the shift produced by a
$Z_2$ translation in one circle of the two-torus. 
In the case of one-quarter of fixed points, $N^{(\rm t)}=4$, it is:
\be
{1 \over 4} \Gamma_{2,2} \ar{0,0}{0,0}+
{1 \over 4} \Gamma_{2,2} \ar{H,0}{G,0}+
{1 \over 4} \Gamma_{2,2} \ar{0,H}{0,G}+
{1 \over 4} \Gamma_{2,2} \ar{H,H}{G,G} \, ,
\ee
where now there are two independent shifts, along the
two independent circles of the two-torus.
In the first case, the function $\Delta^{N^{(\rm t)}}(T,U)
=\Delta^{N^{(\rm t)}}(T)+\Delta^{N^{(\rm t)}}(U)$,
that encodes the $(T,U)$-dependence of the correction on 
the heterotic side, and whose $U$-dependent part gives the
$U$-dependence of the type I corrections, is:
\ba 
\Delta^{8} \left( T,U \right) &=&
-  \log \Im T 
\left\vert  \eta \left(  T    \right) \right\vert^4
-{1 \over 3} \log \Im T
\left\vert\th_4 \left(  T    \right) \right\vert^4\nn \\
&& -  \log \Im U
\left\vert  \eta \left(  U    \right) \right\vert^4
- {1 \over 3} \log \Im U
\left\vert  \th_4 \left(  U    \right) \right\vert^4 \, .
\label{delta8}
\ea 
In the second case, 
\ba
\Delta^{4} \left( T,U \right) &=&
-{1 \over 2} \log \Im T 
\left\vert  \eta \left(  T    \right) \right\vert^4
-{1 \over 2} \log \Im T
\left\vert  \th_4 \left(  T    \right) \right\vert^4 \nn \\
&& -{1 \over 2} \log \Im U 
\left\vert  \eta \left(  U    \right) \right\vert^4
-{1 \over 2} \log \Im U 
\left\vert  \eta \left(  U    \right) \right\vert^4 \, .
\label{delta4}
\ea
The coefficients of the $\log \Im T$, $\log \Im U$
terms are the beta-function coefficients of the gravitational term
as seen on the heterotic side, at the Abelian point:
\be
b_{\rm grav}= { 24 + N^{(\rm t)} - (N_V -N_H) \over 12} \, ,
\ee
where $(N_V-N_H)$, the contribution of the fields originating from the 
currents, is $16$. We stress that $(N_V-N_H)=16$ at both the $U(2)^8$
and $U(1)^{16}$ points. As a consequence, also the functions 
$\Omega \ar{H}{G}$ are the same. 
As we mentioned, this is due to an universality property,
analyzed in Ref.(\cite{kkprn}) for a class of heterotic
models in which the two-torus lattice sum is $\ar{H}{G}$-shifted.
The universality properties there derived, relating models
with the same shift in the lattice sum and the same 
$N_V-N_H$, are valid here, if $N_V$, $N_H$ are the numbers of 
vector and hypermultiplets in the orbifold untwisted sector.

In the $N^{(\rm t)}=16$ case, without the ${\cal D}$-projections,
the result (\ref{rb}) could have been obtained in a more economic way,
without knowing the details of the partition function in the 
$\ar{H}{G}$-twisted sector, by using general properties 
of the amplitudes. In that case, in fact, it is possible to factorize 
the contribution of the two-torus lattice sum, $\Gamma_{2,2} \ar{0}{0}$,
the only piece that contains a dependence on the moduli,
and sum over $(H,G)$ to obtain directly the amplitude 
$\langle R^2_{(\rm o)} \rangle $. 
Owing to the cancellation, in the integrand, of the non-holomorphic pieces
proportional to ${1 \over \Im \tau}$, the latter 
has the general form \cite{kkpr1}:
\be
\langle R^2_{(\rm o)} \rangle =
\int_{\cal F}{ d^2 \tau \over \Im \tau}
\Gamma_{2,2} \ar{0}{0} \left( A \bar{j}(\bar{\tau}) + B \right) \, .
\label{ajb}
\ee 
$A$ and $B$ are constants, and it is intended that the integral has to 
be properly regularized by subtracting the infrared divergence.
In order to conclude for the result, 
it is then enough to observe that the singularity for 
$\tau \to {\rm i} \infty$ in the integrand of (\ref{ajb}), due to the pole 
of the $j$-function, is subtracted by $\langle P^2 \rangle_{(T^2)}$
and that the subtraction of this amplitude, which has a vanishing 
beta-function coefficient, doesn't change 
the constant term $B$, that remains equal to the 
beta-function coefficient of $\langle R^2  \rangle$.  
This implies that $A=0$ and $B=b_{\rm grav}$.

\subsection*{Comparison of the three constructions}

We pass now to the comparison of the $\tilde{\Omega}$ constructions
with the heterotic and type I models. Since 
the $\tilde{\Omega}$ orbifolds match these two string constructions
in a corner of their moduli space, they can be considered as 
particular deformations of the heterotic/type I string.
We start therefore with the analysis in the region of the moduli space
in which these theories match.

The heterotic/type I duality for $Z_2$,
non-freely acting orbifolds was analyzed in Ref. \cite{ap1,ap2},
where it was found that the $T$-dependent, one-loop heterotic contribution
to the effective coupling is mapped, on the type I side, 
into a tree-level plus a non-perturbative contribution.
In the large-$T$ limit, $\Delta^{N^{(\rm t)}}(T)$ behaves as
$ \sim { \pi \over 24} N^{(\rm t)}   \Im T$,
and the heterotic one-loop contribution reproduces 
the term in Eq. (\ref{Ig}) linear in $S^{\prime}$, that parametrizes
the coupling of the gauge fields of the D5-branes sector.
The map is therefore $N^{(\rm t)}\, T \to 3 \times 2^7 \, \pi \, S^{\prime}$.
The coefficient $N^{(\rm t)}$ automatically takes into account
the fact that, when the number of twisted hypermultiplets
is reduced, there is a ``higher level'' realization of the
D5-branes gauge group, reflected in a different strength of the coupling:
$\Im S^{\prime}_{2k}= 2 \Im S^{\prime}_k$, where the level $k$
is related to the rank of the gauge group by $k = 16 \big/ r$.
As usual, the heterotic dilaton--axion field $S$ is dual to the
field $S$ of type I.
In the case of the dual type I orientifolds obtained by T-dualizing
the four coordinates transverse to the D5-branes,
the map is similar, with a simple exchange of the role of the
fields $S$ and $S^{\prime}$.
This is also the expression of the effective coupling in the
$\, \tilde{\Omega} \, $ constructions, for $V_{(4)} \to 0$.
In the opposite limit, $V_{(4)} \to \infty$,
the correction is on the other hand 
approximated by the expression given in Eqs. (\ref{rIIB}).
Matching these constructions with the heterotic theory leads to conclude that
indeed the dominant behavior of the contribution of
the $\tilde{\Omega}$ D9-branes sector to the effective coupling
is given by  (\ref{dtu}), with the following identifications:
\be
T \leftrightarrow S^{\prime} \leftrightarrow \tilde{S}^{-1} \, . 
\ee
At the point in which these theories match with the heterotic string,
we can trade the symmetries in the heterotic field $T$ for
symmetries of the corresponding dual field $\tilde{S}$ in the 
$\tilde{\Omega}$ constructions.
Since these can be seen as deformations 
``toward the bulk'' of type I orientifolds, we argue that 
such symmetries continue to hold also away from the matching point.
For instance, in the orbifold with sixteen fixed points,
in order to understand what happens when we invert the field $\tilde{S}$, 
we can consider to go first to the point at which the theory
matches the heterotic string, and invert there the field 
$S^{\prime} \propto \tilde{S}^{-1}$. On the heterotic side, 
this corresponds to an inversion of the field $T$. Since we are at the 
point at which the two theories match, the invariance of the heterotic 
theory under inversion of $T$ must reflect in an analogous invariance of the
$\tilde{\Omega}$ theory under inversion of $\tilde{S}$.
We can now deform back this theory to a finite $V_{(4)}$:
the new theory must coincide with the one before inversion of 
$\tilde{S}$.  

In the $V_{(4)} \to \infty$ limit, the contribution of the field $S$,
corresponding to the tree level coupling of the D5-branes of the
$\tilde{\Omega}$ orbifolds, is on the other hand missing.
As it was for the cases considered in Ref. \cite{gmono},
also in this case, from the point of view of the
effective theory felt by an observer on a four-dimensional wall 
placed at some point in the bulk, we interpret this decompactification 
as a non-perturbative motion in the field $S$. The absence of a 
contribution linear in this field suggests that  the linear term, 
corresponding to a tree level contribution on the heterotic side, may come 
from the expansion of a function of the type:
\be
\sim \, \log \Im S \vert \th_2 ( S) \vert^4 \, + \,
{\cal O} \left( {\rm e}^{-S} f(T,U)  \right) \, , 
\label{shet}
\ee 
where the second term is a series of terms exponentially suppressed 
both in the large- and small-$S$ limits.
For large $\Im S$, expression (\ref{shet}) reproduces the tree level linear
dependence on this field, while for $S \to 0$, this function
doesn't vanish but diverges only logarithmically in $-1 \big/ S$.
We would therefore be in the presence of a non-perturbative Scherck-Schwarz  
mechanism, providing a phenomenon
analogous to what happens, perturbatively, in the semi-freely acting orbifolds
\footnote{See for instance Ref. \cite{gkr}.}.
We argue therefore that the total correction to the
effective coupling of the $R^2$ term takes the following form: 
\ba
{16 \, \pi^2 \over g^2} & \approx & 16 \, \pi 
\log \Im S \vert \th_2 ( S) \vert^4 \, + 
\Delta^{N^{(\rm t)}}(\tilde{S}^{-1},U)
\,+ \, b \log M_P \big/ \sqrt{p^2} \nn \\
&& \, + \,
{\cal O} \left( {\rm e}^{-S} f(\tilde{S}^{-1},U)  \right) \, . 
\label{full}
\ea
In the above  expression, we must allow for the presence of 
terms, suppressed in both the large- and small-$S$ limits,
that mix the contributions of the three moduli, $S$, $\tilde{S}$, $U$.

\subsection*{\it The gauge couplings}

The analysis of the effective coupling of the $R^2$ term
allows us to learn something also about the effective couplings
of the gauge group in the $\tilde{\Omega}$ constructions. 
From the bulk point of view, the states living on the 
D5-branes, appear as massive. 
This is true not only at the Abelian point,
but also for configurations with a larger gauge group, such as those
one can obtain by considering Dp-branes, p $>$ 5, via T-dualization of
some  coordinates and switching-off of some Wilson lines.
These are the states corresponding to the heterotic perturbative gauge group.
The $\tilde{\Omega}$ constructions represent therefore non-perturbative 
deformations of the heterotic string in which, due to
a non-perturbative Higgs phenomenon, all the states of the ``currents''
acquire a mass, depending on the field $S$.
The states of the D9-branes of the $\tilde{\Omega}$ constructions,
on the other hand, are generically massless (we are not considering here
possible mass terms required by anomaly cancellation at 
the contact point with the heterotic/type I theory).
The only massless gauge bosons of these constructions
originate from this sector,
with coupling parametrized by the field $S^{\prime}$;
the analysis of the previous section tells us also that
the effective coupling is essentially invariant under inversion of this 
field.
We expect therefore, for the orbifold with
maximal number of fixed points, a behavior roughly of the type:
\be
{1 \over \alpha} \, \approx \, 
\log \Im \tilde{S} | \eta ( \tilde{S} ) |^4 \, .
\label{a5np}
\ee
(In the models with a reduced rank of the gauge group,
the approximate behavior is obtained  instead from the  expressions
(\ref{delta8}), (\ref{delta4})). 
At a generic point in the moduli space of this theory,
however, the gauge beta function is not vanishing,
and this expression may receive
corrections that depend, among  the others, also
on the field $S$, that parametrizes the couplings of the D5-branes sector;
this implies that this sector ``feels'' the presence of the other one.
Expression (\ref{a5np}) then simply
gives the dominant behavior of the coupling, 
as a function of one modulus, indicating the existence of an
(approximate?) S-duality. The coupling of the (massive) states of the
other sector behaves on the other hand as:
\be 
{1 \over \tilde{\alpha}} \approx \log \Im S | \vartheta ( S) |^4 \, ,
\label{a9np}
\ee
and the mass of these states scales as the inverse of the field $S$.
In the limit $S \to \infty$, corresponding to the heterotic weak coupling,
these states become massless, and the gauge group gets enhanced:
at this point, there can be a transition to an heterotic/type I phase.

\noindent

\vskip 0.3cm
\setcounter{section}{6}
\setcounter{equation}{0}
\section*{\normalsize{\bf 6. Conclusions}}

In this work we have analyzed the bulk theory of a class of 
$Z_2$ orbifolds of the type IIB string,
compactified to four space-time dimensions, in the presence of D5-branes.
Cancellation of the RR charge fixes uniquely the number of D5-branes.
The $Z_2$ orbifold projection, parallel to the D5-branes, 
adds to the spectrum the states of the orbifold fixed points, that appear 
as D9-branes. From an observer sitting on a four-dimensional probe
located at a point in the bulk, supersymmetry appears broken 
to ${\cal N}_4=2$ even outside of the D5-branes. This is due to the
interaction of branes and bulk states.
In the limit of compactification to zero size of the space transverse
to the D5-branes, the effective theory is expected to match with
that of an orbifold of the type I string (and therefore, by duality, 
also heterotic string), compactified to four dimensions. On the other hand,
when the transverse space is decompactified, we expect that,
in the case in which there is a local cancellation of the 
RR charge between D5 branes and corresponding orientifold planes,
from the bulk point of view the D5-branes effects appear to be suppressed.
This however does not necessarily lead to a restoration of a larger
amount of supersymmetry in the bulk: when the orbifold projection
has fixed points, there is indeed no point in the moduli space of these
constructions in which there is such a kind of phenomenon.
We have considered the effective four dimensional action
and the corrections to the coupling of the $R^2$ term.
By matching them,
at a corner in their moduli space, with heterotic orbifolds,
we interpreted these constructions as providing a 
non-perturbative deformations of the heterotic/type I string.
The motion toward a decompactification regime can then be seen,
from the point of view of the four dimensional effective theory, as
a non-perturbative motion toward a particular strong coupling regime.
This interpretation is
somehow reminiscent of the interpretation of the decompactification
of the eleventh coordinate of the M-theory as a motion
toward a strong string coupling regime. 
Indeed, we believe that these constructions, as well as those
considered in Ref. \cite{gmono}, provide
effective realizations of Scherck-Schwarz mechanisms that 
lift the mass of the heterotic perturbative currents; they probably
have their appropriate realization as
(semi)freely acting orbifolds of higher dimensional theories,
a hint of whom can be found in Refs. \cite{k}--\cite{b}.
In this framework, the constructions in which the orbifold projection 
acts freely would then
realize phases in which all the gauge bosons are massive, while
those in which the projections acts only semi-freely have 
massless gauge bosons associated to the orbifold fixed points.
Such bosons can be entirely non-perturbative from the heterotic
point of view: this is indeed the case considered in this work,
and is realized
when they are associated to the fixed points of a projection 
non-perturbative from the point of view of the heterotic string.

\vskip 1.cm
\centerline{\bf Acknowledgements}
\noindent
I thank C. Angelantonj, I. Antoniadis,
C. Bachas, K. Benakli, R. Blumenhagen, D. L\"{u}st, P. Mayr, W. Nahm,
Y. Oz and A. Sagnotti for valuable discussions, the
\'Ecole Normale of Paris and the CERN Theory Division for hospitality.

\vspace{1.5cm}


\end{document}